\documentclass[aps,prl,twocolumn,superscriptaddress,calc,epsfig,10pt]{revtex4-1}
\usepackage{graphicx}
\usepackage{graphics}
\usepackage{amssymb,amsmath}
\usepackage{subcaption}

\usepackage{wasysym}

\usepackage{color}
\usepackage[normalem]{ulem}

\newcommand{\mitCUAaddress}{Department of Physics, MIT-Harvard Center for Ultracold Atoms, and Research Laboratory of Electronics, MIT, Cambridge, Massachusetts 02139, USA}

\newcommand{\ket}{\right\rangle}

\newcommand{\bra}{\left\langle}

\begin{document}

\title{Observation of 2D fermionic Mott insulators of $^{40}$K with single-site resolution}

\author{Lawrence W. Cheuk, Matthew A. Nichols, Katherine R. Lawrence, Melih Okan, Hao Zhang, and Martin W. Zwierlein}
\date{\today}

\affiliation{\mitCUAaddress}

\begin{abstract}
We report on the site-resolved observation of characteristic states of the two-dimensional repulsive Fermi-Hubbard model, using ultracold $^{40}$K atoms in an optical lattice. By varying the tunneling, interaction strength, and external confinement, we realize metallic, Mott-insulating, and band-insulating states. We directly measure the local moment, which quantifies the degree of on-site magnetization, as a function of temperature and chemical potential. Entropies per particle as low as $0.99(6)\,k_B$ indicate that nearest-neighbor antiferromagnetic correlations should be detectable using spin-sensitive imaging.

\end{abstract}
\maketitle

Strongly correlated fermions present a fundamental challenge to many-body physics, as no general method exists to predict what phenomena will emerge~\cite{Troyer2005}. Ultracold gases of fermionic atoms have shown promise as a clean, highly controllable platform for studying such systems~\cite{ingu08varenna,bloc08review}. 
One prominent example is the realization of strongly coupled fermionic superfluids, enabled by the enhanced interactions that arise near a Feshbach resonance~\cite{Zwerger2011BECBCS,Zwierlein2014NovelSuperfluids}. 
Another class of strongly correlated systems well-suited for simulation with ultracold atoms is lattice models, in which the kinetic and interaction energies can be set to comparable strengths~\cite{bloc08review,Esslinger2010FermiHubbard}. One such model is the Fermi-Hubbard model, believed to capture the essential aspects of high-temperature superconductivity~\cite{ande87,lee06hightc}.

The realization of the Fermi-Hubbard model at low entropies has been a longstanding goal in ultracold atom experiments.
Mott-insulating behavior has been observed in three dimensions (3D) via reduction of double occupancies and compressibility~\cite{jord08,schn08,taie2012,duarte2015}. Short-range antiferromagnetic correlations above the N\'eel temperature were observed via Bragg scattering and dimerized lattices \cite{Greif2013Magnetism,Hart2015FermiHubbard,Greif2015}. Recently, the equation of state of the Fermi-Hubbard model has been measured in two dimensions (2D) for spin-1/2 and in 3D for higher spin values ~\cite{Cocchi2015,hofrichter2015}. However, these experiments relied on conventional imaging techniques that do not allow site-resolved measurements of microscopic quantities. 

Such microscopic measurements first became possible in bosonic systems through the development of quantum gas microscopes with single-site resolution, and have enabled studies of ordering, spatial structures, and correlations in the Bose-Hubbard model ~\cite{bakr2010MottInsulator,sherson2010microscope,Endres2011stringorder,Cheneau2012Lightcone}. Recently, the ability to perform single-site imaging has been extended to the two workhorse fermionic isotopes of alkali atoms, $^{6}$Li and $^{40}$K ~\cite{Cheuk2015,Haller2015,Parsons2015,Edge2015,Omran2015}. While $^{6}$Li has faster lattice dynamics due to its smaller mass, $^{40}$K features a larger fine structure splitting, which is beneficial for implementing spin-dependent potentials and spin-orbit coupling.

After initial demonstrations of site-resolved imaging of non-degenerate Fermi gases, the goal has been to apply these imaging techniques to low-entropy degenerate gases in order to study quantum many-body phenomena. Within the past few months, Pauli blocking was directly observed in a spin-polarized gas of $^{6}$Li \cite{Omran2015}, and the metallic, Mott-insulating and band-insulating states of the 2D Fermi-Hubbard model have been directly detected, both in $^{6}$Li \cite{Greif2016}, and, as reported in this paper, in $^{40}$K. In this work, we also demonstrate the formation of local moments at half-filling as the temperature is lowered.
\begin{figure*}[t]
\centering
\includegraphics[scale=1.0]{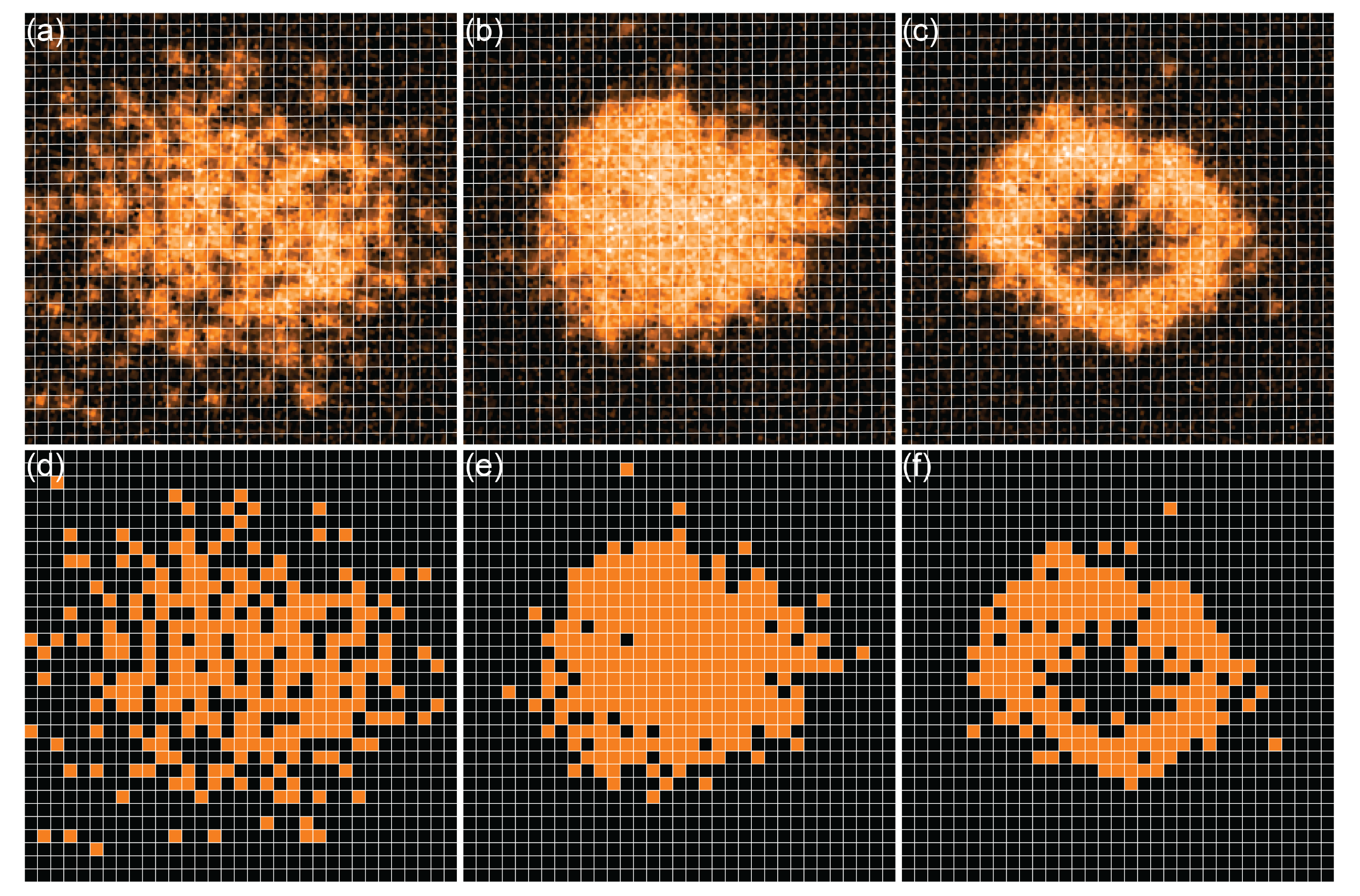}
\caption{Metallic, Mott-insulating, and band-insulating states under the quantum gas microscope. Observed fluorescence images, showing (a) the metallic state, with $\mu_0/h = 280(40)\,\rm{Hz}$, $\omega = 2\pi \times 111(3)\,\rm{Hz}$, and $U/8\bar{t} =0.33(4)$ with $U/h = 540(60)\,\rm{Hz}$; (b) the Mott-insulating state, with $\mu_0/h=624(22)\,\rm{Hz} $, $\omega = 2\pi \times 115(3)\,\rm{Hz}$, and $U/8\bar{t} = 12.3(8)$ with $U/h = 1350(50)\,\rm{Hz}$; (c) the band-insulating state, with $\mu_0/h=1450(40)\,\rm{Hz}$, $\omega = 2\pi \times 181(3)\,\rm{Hz}$, and $U/8\bar{t} = 2.6(1)$ with $U/h = 1007(40)\,\rm{Hz}$. (d,e,f) Reconstructed detected site occupations corresponding to (a,b,c), respectively.}

\label{fig1}
\end{figure*}

\begin{figure*}[t]
\centering

\includegraphics[scale=1.0]{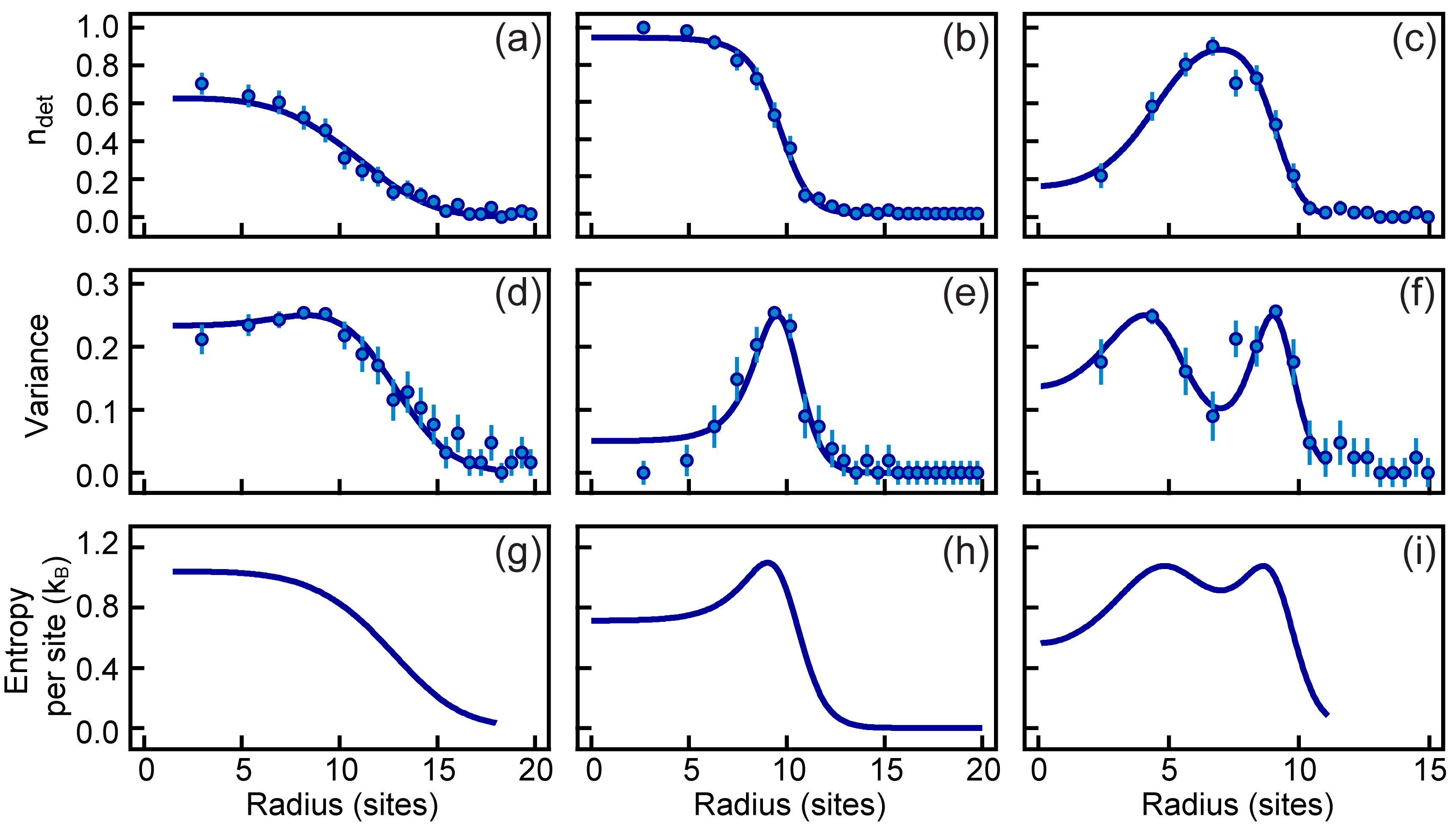}
\caption{Radially averaged detected site occupation (a,b,c), variance (d,e,f), and entropy (g,h,i), with theoretical curves. (a,d,g) Metallic state, with $\mu_0/h= 280(40)\,\rm{Hz}$ and $k_BT/U = 1.46(18)$; average entropy per particle $S/N=1.7(1)\,k_B$. (b,e,h) Mott-insulating central region, with $\mu_0/h= 624(22)\,\rm{Hz}$ and $k_BT/U = 0.09(1)$; $S/N=1.23(6)\,k_B$. (c,f,i) Band-insulating center and Mott-insulating annular region, with $\mu_0/h=1450(40)\,\rm{Hz}$ and $k_BT/U = 0.18(2)$; $S/N=0.99(6)\,k_B$. The profiles were fitted to NLCE data with $U/\bar{t}=3$ for (a,d,g) and to HTSE for (b,e,h). For (c,f,i), profiles were fitted to NLCE data with $U/\bar{t}=21$, shown in solid.}
\label{fig2}
\end{figure*}

\begin{figure*}[t]
\centering
\includegraphics[scale=1.0]{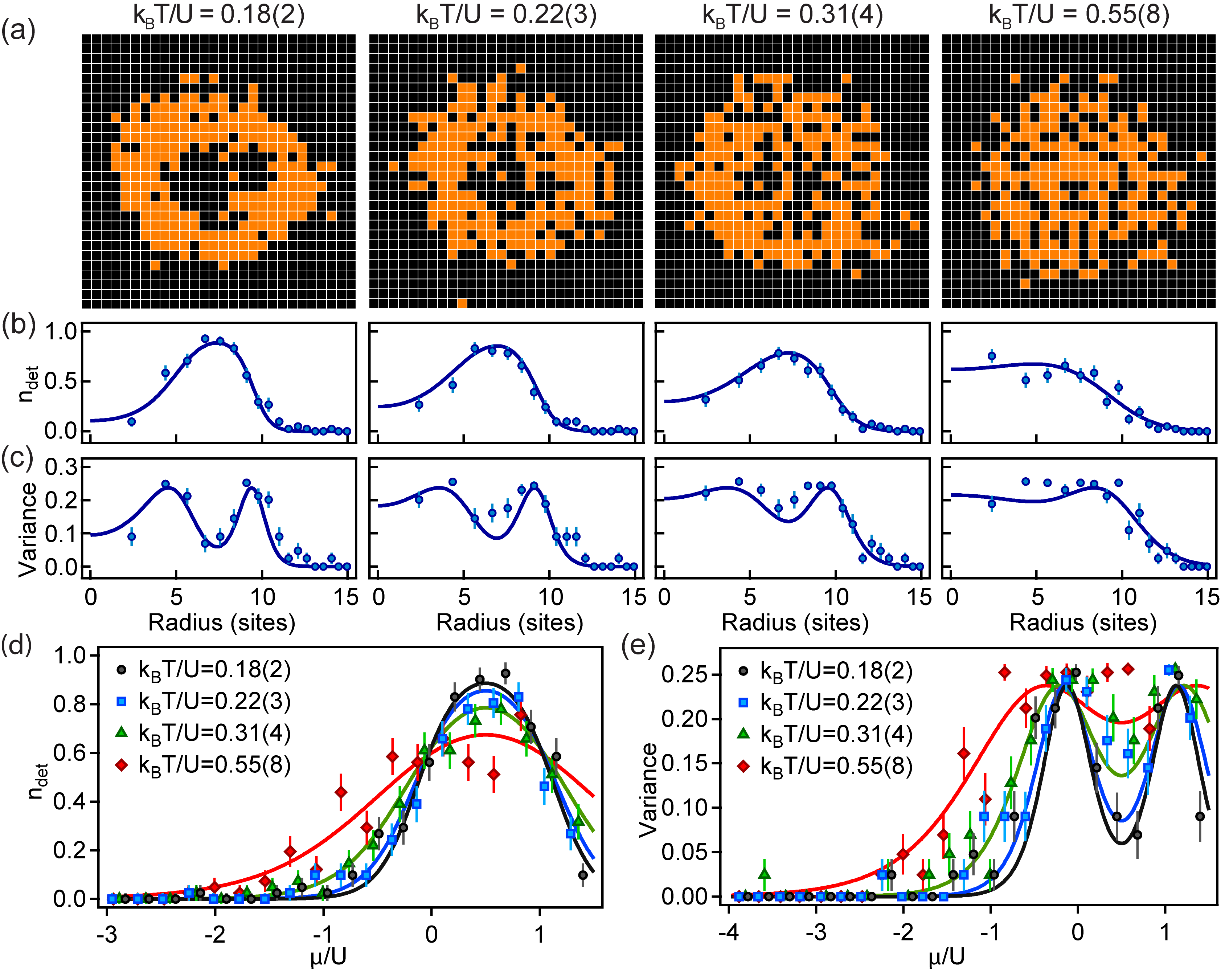}
\caption{Heating of Mott and band insulators.  (a) Site-resolved images, (b) density profiles, and (c) variances for temperatures $k_B T/U = 0.18(2),0.22(3),0.31(4),0.55(8)$ (left to right) at fixed $U/8\bar{t} = 2.6(1)$ and $\omega = 2\pi \times 183(3)\,\rm{Hz}$, with fitted curves from HTSE (solid). (d,e) Radially averaged observed filling and variance, respectively, for all four temperature values as a function of chemical potential, calculated from the fitted global chemical potential.}
\label{fig3}
\end{figure*}

\begin{figure}[t]
\centering
\includegraphics[scale=1.0]{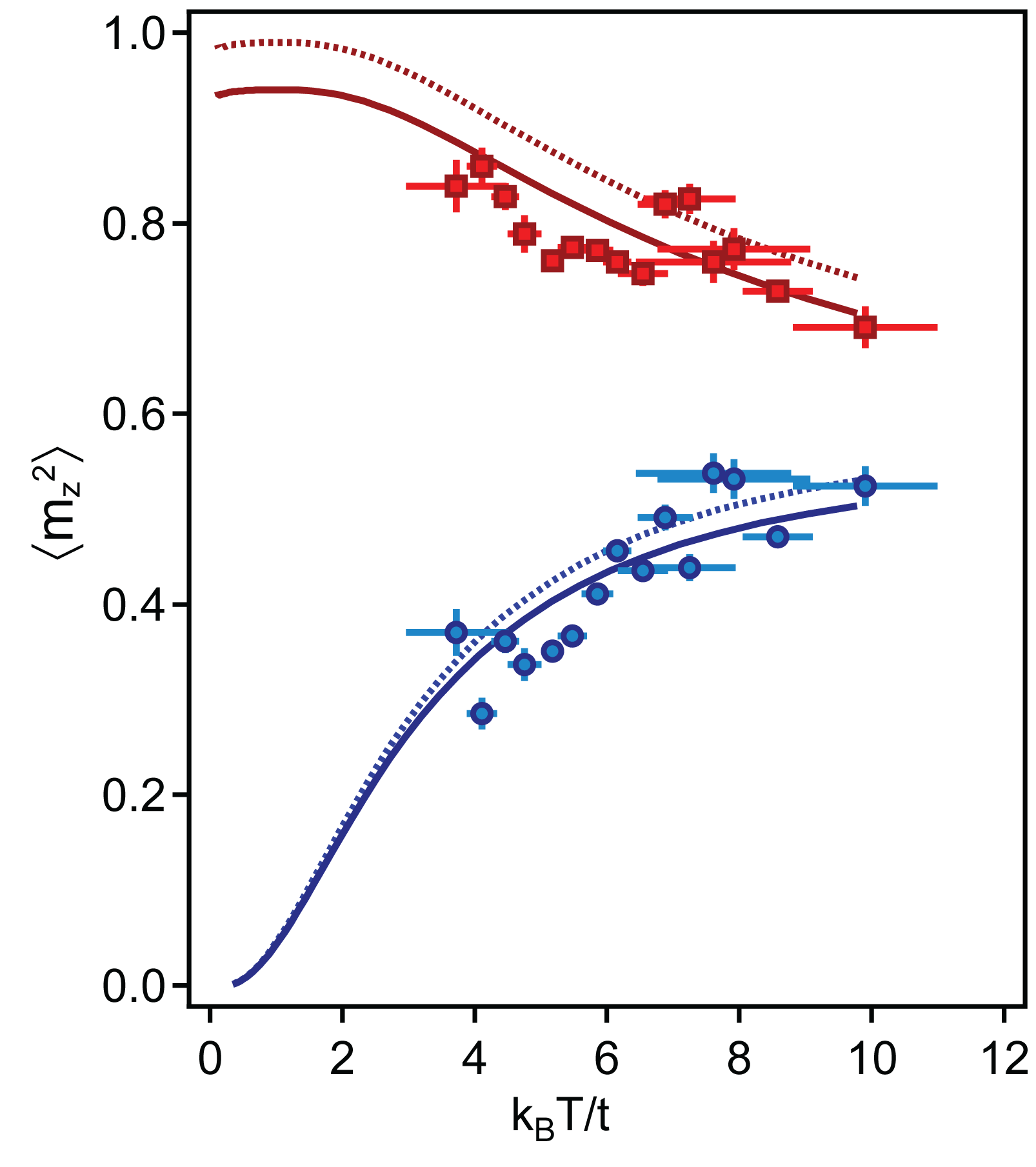}
\caption{Local moment $\langle m_z^2 \rangle$ as a function of temperature at $U/8\bar{t} = 2.6(1)$ for $\mu/U = 0.5$ (red squares) and $\mu/U = -0.25$ (blue circles). The local moments are extracted from 117 samples. For each sample, the temperature and global chemical potential are determined by fitting to HTSE in the outer regions where $n_{\text{det}}<0.25$. NLCE data at $U/t=21$ with and without adjustment for imaging fidelity are shown in solid and dotted lines, respectively.}
\label{fig4}
\end{figure}

Our system is described by the single-band 2D Hubbard Hamiltonian with two spin states on a square lattice,
\begin{equation*}
\hat{H}= -t\hspace{-0.1cm}\sum_{\bra i,j\ket,\sigma} \hspace{-0.1cm} (\hat{c}_{i\sigma}\hat{c}^\dag_{j\sigma} + \text{h.c.}) + U \sum_i \hat{n}_{i \uparrow} \hat{n}_{i \downarrow} + \sum_{i,\sigma} (V_i-\mu_0) \hat{n}_{i\sigma}
\end{equation*}
where $\hat{c}_{i\sigma}$ ($\hat{c}_{i\sigma}^\dagger$) is the fermion annihilation (creation) operator for spin $\sigma=\{\uparrow,\downarrow\}$ on site $i$, $n_{i\sigma} =  \hat{c}_{i,\sigma}^\dagger \hat{c}_{i,\sigma}$ is the number operator on site $i$, and angle brackets indicate summation over nearest neighbors. $U$ and $t$ denote the on-site interaction energy and nearest-neighbor hopping amplitude, respectively, while $\mu_0$ is the chemical potential and $V_i$ is the on-site energy due to the overall trapping potential. The trapping potential is approximated by $V_i = \frac{1}{2} m \omega^2 d_i^2 a^2$, where $m$ is the atomic mass, $\omega$ is the global trapping frequency, $d_i$ is the distance in lattice sites from the center of the trap, and $a$ is the lattice spacing. 

Despite the simplicity of the Hamiltonian, this model is theoretically intractable and has been solved only in special cases. At weak interactions ($U/8t < 1$) or when the average filling is well below unity, the system is metallic. If the chemical potential is high enough to fill all available states, the system becomes a band insulator, with two opposite-spin atoms per site. At strong interactions ($U/8t \gg 1$) and at half-filling, another insulating state, the Mott insulator, appears when the temperature $k_B T\ll U$. At temperatures well below the super-exchange scale of $4t^2/U$, long-range antiferromagnetic correlations arise. It is conjectured that $d$-wave superconductivity emerges upon doping a magnetically ordered Mott insulator~\cite{ande87,lee06hightc}. Within the local density approximation (LDA), the overall harmonic confining potential leads to a spatially varying local chemical potential, and thus metallic, Mott-insulating, and band-insulating states can coexist within the same sample ~\cite{Scarola2009, Paiva2010}.

To realize this model, we begin by sympathetically cooling $^{40}$K atoms with $^{23}$Na atoms in a magnetic trap. The $^{40}$K atoms are then transferred into an optical dipole trap, and an equal mixture of hyperfine states $\left|F=9/2, m_F=-9/2\ket$ and $\left| 9/2, -7/2 \ket$ is created. After evaporation and transport, we obtain a highly oblate layer of $\sim 300$ $^{40}$K atoms in the $x$-$y$ plane $7\,\mu\rm{m}$ underneath the imaging system. Subsequently, we ramp up a square optical lattice in the $x$-$y$ plane, with lattice spacing $a=541\,\rm{nm}$, to a depth of either $6\,E_R$, $12\,E_R$, or $18\,E_R$, where $E_R = \frac{\hbar^2}{2m} \left(\frac{\pi}{a}\right)^2$. The laser beams that create the $x$-$y$ lattice also interfere to form a lattice along $z$ with 3~$\mu$m spacing, where only one layer is populated. We use the lattice depth to tune the Hubbard parameters $t$ and $U$, without utilizing any Feshbach resonances. For this work, the magnetic field is set to $4.5$~G, where the scattering length is 170~$a_0$, $a_0$ being the Bohr radius. While the lattice is ramped up, the radial confinement within the plane is brought to the desired value. For imaging, the lattice depth is quickly increased to $\sim 1000\,E_R$, while an additional lattice along the $z$-direction with spacing $532\,\rm{nm}$ is also applied. 

We detect the occupation on each lattice site using Raman sideband cooling, which cools the atoms while  scattering enough photons to produce a fluorescence image~\cite{Cheuk2015}. This imaging technique, combined with an image reconstruction algorithm, allows us to determine the occupation of a given lattice site with a measured imaging fidelity of $95\%$. Because pairs of atoms residing on the same site are lost during imaging due to light-assisted collisions \cite{depu99}, only the parity of the occupation is detected. Additionally, this imaging method does not distinguish between the two spin states. 
The average detected occupation at site $i$ is thus given by $n_{\text{det}}(i) = \bra \hat{n}_{\text{det}}(i)\ket$, where $\hat{n}_{\text{det}}(i) = \hat{n}_{i\uparrow}+ \hat{n}_{i\downarrow}- 2\hat{n}_{i\uparrow}\hat{n}_{i\downarrow}$.

We directly observe the metallic, Mott-insulating, and band-insulating states using three configurations of lattice depths and radial confinements. The three different samples are prepared identically until the 2D lattice ramp, where both the depth of the lattice and the radial confinement are adjusted.
In Fig.~1 we show the site-resolved fluorescence images and the reconstructed detected site occupations. In Fig.~2 we show profiles of the corresponding radially averaged parity-projected densities $n_{\text{det}}$ and their variances. The Mott-insulating and band-insulating states are both expected to show suppressed variance in $n_{\text{det}}$. In particular, the variance is suppressed in Mott-insulating regions due to the charge gap, which is $U$ at half-filling; in the band-insulating regions, it is suppressed instead by Pauli blocking. In the metallic regions the variance is not suppressed, and in the case of half-filling it equals 0.25, since a site is equally likely to be empty, doubly occupied, or singly occupied by an atom of either spin state. The variance can either be directly measured, or obtained via $\bra \hat{n}^2_{\text{det}} \ket - \bra \hat{n}_{\text{det}} \ket^2 = n_{\text{det}} (1-n_{\text{det}})$. This is due to the operator identity $\hat{n}_{i \sigma}^2  =  \hat{n}_{i\sigma}$ for fermions, which implies $\bra \hat{n}^2_{\text{det}} \ket = \bra \hat{n}_{\text{det}} \ket$, and more generally all moments of $n_{\text{det}}$ can be found from $n_{\text{det}}$ itself.

The metallic state, with peak occupation $0.7$ and peak variance $\sim 0.25$, is shown in Fig. 1(a,d) and 2(a,d). Here the lattice depth is $6\,E_R$ and the radial confinement is $\omega=2\pi \times 111(3)\,\rm{Hz}$. This corresponds to $U/8\bar{t}=0.33(4)$, where  $\bar{t}=\sqrt{t_x t_y}$ is the mean hopping amplitude, with $t_x$ ($t_y$) the mean hopping amplitude along the $x$ ($y$) direction. In order to observe the Mott insulator, shown in Fig. 1(b,e) and 2(b,e), we increase the interaction to $U/8\bar{t}=12.3(8)$ by increasing the lattice depth to $18\,E_R$ and the trap confinement to $\omega = 2\pi \times 115(3)\,\rm{Hz}$. The detected site occupation flattens to $0.98(2)$ at the trap center, with a corresponding variance less than $0.03$. To observe the band insulator, shown in Fig. 1(c,f) and 2(c,f), we increase the global chemical potential, by increasing the trap confinement to $\omega=2\pi \times 181(3) \,\rm{Hz}$, while reducing the interaction to $U/8\bar{t}=2.6(1)$, by lowering the lattice depth to $12\,E_R$.  At the center the detected density is depleted and the variance is suppressed, indicating a band-insulating region with two atoms per site. Due to the varying local chemical potential across the trap, a surrounding Mott-insulating annular region is also visible. The metallic regions that border the insulating regions are clearly evidenced by the increased variance. 

To characterize the atomic clouds, we fit the radially averaged parity-projected density $n_{\text{det}}$ to the equation of state of the spin-balanced Hubbard model obtained either through numerical linked cluster expansion (NLCE) data~\cite{khatami2011}, for $U/8\bar{t} =0.33(4),\,2.6(1)$, or from the high temperature series expansion (HTSE) in $\bar{t}/{k_BT}$~\cite{Oitmaa2006}, for $U/8\bar{t}=12.3(8)$. From these fits, we extract temperatures of $k_BT/U = 0.55(9),\, 0.09(1), \, 0.18(2)$ for the three configurations shown in Fig.~2(a,d,g), (b,e,h), and (c,f,i), respectively. From the fits, we deduce the local entropy per site, shown in Fig.~2(g,h,i), and the trap-averaged entropy per particle. These curves illustrate redistribution of entropy between the different regions of the trap. There is a local reduction of entropy in the Mott and band-insulating regions, with a corresponding increase of entropy in the metallic rings. Additionally, we observe that the average entropy per particle is 1.7(1)$k_B$, 1.23(6)$k_B$, and 0.99(6)$k_B$ for the three configurations.

In order to explore the effects of temperature, we heat samples at $U/8\bar{t} = 2.6(1)$ and confinement of $\omega = 2\pi \times 181(3)\,\rm{Hz}$ by varying the hold time in the lattice up to 3\,s. In Fig.~3(a), we show the reconstructed site occupations for four temperatures from $k_BT/U = 0.18(2)$ to $0.55(8)$. As the temperature increases, singly-occupied sites are created in the band-insulating region as $k_BT$ approaches $\mu_0$, while double occupancies and holes appear in the Mott-insulating region as $k_BT$ approaches $U$. The radially averaged density profiles, shown in Fig.~3(b), are fitted with HTSE to extract the temperature and chemical potential. In Fig.~3(c), we show the measured variance for the samples from Fig.~3(a). The variance is suppressed in insulating regions at low temperatures, but approaches 0.25 throughout the sample at high temperatures. To extract trap-independent properties, we use the fitted value of $\mu_0$ and the trap frequency $\omega$ to determine the local chemical potential $\mu = \mu_0 - \frac{1}{2}m\omega^2 d_i^2 a^2 $. Under the LDA, the local properties are equivalent to those of a homogeneous system at the same chemical potential. Radial profiles can then be converted to profiles with varying $\mu/U$, as shown in Fig.~3(d) and (e) for the site occupation and variance, respectively.

While the detected site occupation $n_{\text{det}}$ does not allow one to obtain the total density $\bra \hat{n}_{\downarrow}+ \hat{n}_{\uparrow}\ket$ or the double occupancy $\bra \hat{n}_{\downarrow} \hat{n}_{\uparrow}\ket$ separately, it directly gives the local moment $\bra m_z^2 \ket = \bra (\hat{n}_{\uparrow} - \hat{n}_{\downarrow})^2\ket =\bra \hat{n}_{\uparrow} + \hat{n}_{\downarrow} - 2 \hat{n}_{\uparrow} \hat{n}_{\downarrow}  \ket =  \bra \hat{n}_{\text{det}} \ket$\,\cite{Hirsch1985}. In the strong coupling limit $U\gg t$ and at half-filling, as the temperature is lowered below $\sim U$, the local moment is expected to approach unity as the system enters the Mott-insulating state. At even lower temperatures, near the super-exchange scale $t^2/U$, the moment is expected to slightly decrease, signaling reduced localization as magnetic interactions become important\,\cite{Paiva2001}. The moment, directly given by $n_{\text{det}}$, can thus show signatures of super-exchange, albeit at temperatures lower than those accessed in the current work. In Fig.~4, we show the local moment at half-filling ($\mu=U/2$) as a function of temperatures for the same parameters as in Fig.~3. To determine the half-filling point, the detected occupation is fitted to HTSE in the outer regions of the sample where $n_{\text{det}} <0.25$, from which we extract the temperature and global chemical potential $\mu_0$. Note that at half-filling, a measurement of the local moment also yields the double occupancy via $\langle \hat{n}_{\uparrow} \hat{n}_{\downarrow}\rangle = (1-\langle m_z^2\rangle)/2$. We also show the measured temperature dependence of the moment at $\mu=-U/4$. Note that $\bra m_z^2\ket$ is symmetric about $\mu=U/2$, a consequence of the particle-hole symmetry of the Fermi-Hubbard model on a bipartite lattice. Thus the behavior of the moment versus temperature at $\mu=-U/4$ is representative of the metallic regions both below and above half-filling. After correction for imaging fidelity, the data for both values of $\mu$ are consistent with the NLCE predictions. 

In summary, we have directly observed with single-site resolution the Mott-insulating, band-insulating, and metallic states of the 2D Hubbard model using fermonic $^{40}$K in an optical lattice. We measure entropies as low as $0.99(6)\,k_B$ per particle, indicating that short-range antiferromagnetic spin correlations should be present~\cite{khatami2011,LeBlanc2013,Chiesa2011}. The Mott insulator provides a well-controlled initial state for further studies, such as the properties of one-dimensional Hubbard chains and dynamics of magnetic polarons \cite{lee06hightc,weitenberg2011}. Additionally, the presence of $^{23}$Na in our system, combined with the recently demonstrated creation of ground state $^{23}$Na$^{40}$K molecules~\cite{park2015groundstate}, opens the possibility to study lattice models with long-range and anisotropic interactions at the single-site level.

\begin{acknowledgments}
The authors would like to thank David Reens and Thomas Lompe for experimental support, and Zoran Hadzibabic for a critical reading of the manuscript. This work was supported by the NSF, AFOSR-PECASE, AFOSR-MURI on Exotic Phases of Matter, ARO-MURI on Atomtronics, a grant from the Army Research Office with funding from the DARPA OLE program, and the David and Lucile Packard Foundation. KRL was supported by the Fannie and John Hertz Foundation and the NSF GRFP. MAN was supported by the DoD through the NDSEG Fellowship Program.
\end{acknowledgments}

\bibliographystyle{apsrev4-1}

\begin{thebibliography}{38}%
\makeatletter
\providecommand \@ifxundefined [1]{%
 \@ifx{#1\undefined}
}%
\providecommand \@ifnum [1]{%
 \ifnum #1\expandafter \@firstoftwo
 \else \expandafter \@secondoftwo
 \fi
}%
\providecommand \@ifx [1]{%
 \ifx #1\expandafter \@firstoftwo
 \else \expandafter \@secondoftwo
 \fi
}%
\providecommand \natexlab [1]{#1}%
\providecommand \enquote  [1]{``#1''}%
\providecommand \bibnamefont  [1]{#1}%
\providecommand \bibfnamefont [1]{#1}%
\providecommand \citenamefont [1]{#1}%
\providecommand \href@noop [0]{\@secondoftwo}%
\providecommand \href [0]{\begingroup \@sanitize@url \@href}%
\providecommand \@href[1]{\@@startlink{#1}\@@href}%
\providecommand \@@href[1]{\endgroup#1\@@endlink}%
\providecommand \@sanitize@url [0]{\catcode `\\12\catcode `\$12\catcode
  `\&12\catcode `\#12\catcode `\^12\catcode `\_12\catcode `\%12\relax}%
\providecommand \@@startlink[1]{}%
\providecommand \@@endlink[0]{}%
\providecommand \url  [0]{\begingroup\@sanitize@url \@url }%
\providecommand \@url [1]{\endgroup\@href {#1}{\urlprefix }}%
\providecommand \urlprefix  [0]{URL }%
\providecommand \Eprint [0]{\href }%
\providecommand \doibase [0]{http://dx.doi.org/}%
\providecommand \selectlanguage [0]{\@gobble}%
\providecommand \bibinfo  [0]{\@secondoftwo}%
\providecommand \bibfield  [0]{\@secondoftwo}%
\providecommand \translation [1]{[#1]}%
\providecommand \BibitemOpen [0]{}%
\providecommand \bibitemStop [0]{}%
\providecommand \bibitemNoStop [0]{.\EOS\space}%
\providecommand \EOS [0]{\spacefactor3000\relax}%
\providecommand \BibitemShut  [1]{\csname bibitem#1\endcsname}%
\let\auto@bib@innerbib\@empty
%</preamble>
\bibitem [{\citenamefont {Troyer}\ and\ \citenamefont
  {Wiese}(2005)}]{Troyer2005}%
  \BibitemOpen
  \bibfield  {author} {\bibinfo {author} {\bibfnamefont {M.}~\bibnamefont
  {Troyer}}\ and\ \bibinfo {author} {\bibfnamefont {U.-J.}\ \bibnamefont
  {Wiese}},\ }\href@noop {} {\bibfield  {journal} {\bibinfo  {journal} {Phys.
  Rev. Lett.}\ }\textbf {\bibinfo {volume} {94}},\ \bibinfo {pages} {170201}
  (\bibinfo {year} {2005})}\BibitemShut {NoStop}%
\bibitem [{\citenamefont {Inguscio}\ \emph {et~al.}(2008)\citenamefont
  {Inguscio}, \citenamefont {Ketterle},\ and\ \citenamefont
  {Salomon}}]{ingu08varenna}%
  \BibitemOpen
  \bibinfo {editor} {\bibfnamefont {M.}~\bibnamefont {Inguscio}}, \bibinfo
  {editor} {\bibfnamefont {W.}~\bibnamefont {Ketterle}}, \ and\ \bibinfo
  {editor} {\bibfnamefont {C.}~\bibnamefont {Salomon}},\ eds.,\ \href@noop {}
  {\emph {\bibinfo {title} {Ultracold Fermi Gases}}},\ Proceedings of the
  International School of Physics "Enrico Fermi", Course CLXIV, Varenna, 20 -
  30 June 2006\ (\bibinfo  {publisher} {IOS Press, Amsterdam},\ \bibinfo {year}
  {2008})\BibitemShut {NoStop}%
\bibitem [{\citenamefont {Bloch}\ \emph {et~al.}(2008)\citenamefont {Bloch},
  \citenamefont {Dalibard},\ and\ \citenamefont {Zwerger}}]{bloc08review}%
  \BibitemOpen
  \bibfield  {author} {\bibinfo {author} {\bibfnamefont {I.}~\bibnamefont
  {Bloch}}, \bibinfo {author} {\bibfnamefont {J.}~\bibnamefont {Dalibard}}, \
  and\ \bibinfo {author} {\bibfnamefont {W.}~\bibnamefont {Zwerger}},\
  }\href@noop {} {\bibfield  {journal} {\bibinfo  {journal} {Rev. Mod. Phys.}\
  }\textbf {\bibinfo {volume} {80}},\ \bibinfo {pages} {885} (\bibinfo {year}
  {2008})}\BibitemShut {NoStop}%
\bibitem [{\citenamefont {Zwerger}(2011)}]{Zwerger2011BECBCS}%
  \BibitemOpen
  \bibinfo {editor} {\bibfnamefont {W.}~\bibnamefont {Zwerger}},\ ed.,\
  \href@noop {} {\emph {\bibinfo {title} {The BCS-BEC crossover and the unitary
  {Fermi} gas}}},\ Vol.\ \bibinfo {volume} {836}\ (\bibinfo  {publisher}
  {Springer},\ \bibinfo {year} {2011})\BibitemShut {NoStop}%
\bibitem [{\citenamefont {Zwierlein}(2014)}]{Zwierlein2014NovelSuperfluids}%
  \BibitemOpen
  \bibfield  {author} {\bibinfo {author} {\bibfnamefont {M.~W.}\ \bibnamefont
  {Zwierlein}},\ }in\ \href@noop {} {\emph {\bibinfo {booktitle} {Novel
  Superfluids, Vol. 2}}},\ \bibinfo {editor} {edited by\ \bibinfo {editor}
  {\bibfnamefont {K.-H.}\ \bibnamefont {Bennemann}}\ and\ \bibinfo {editor}
  {\bibfnamefont {J.~B.}\ \bibnamefont {Ketterson}}}\ (\bibinfo  {publisher}
  {Oxford University Press},\ \bibinfo {address} {Oxford},\ \bibinfo {year}
  {2014})\BibitemShut {NoStop}%
\bibitem [{\citenamefont {Esslinger}(2010)}]{Esslinger2010FermiHubbard}%
  \BibitemOpen
  \bibfield  {author} {\bibinfo {author} {\bibfnamefont {T.}~\bibnamefont
  {Esslinger}},\ }\href@noop {} {\bibfield  {journal} {\bibinfo  {journal}
  {Annual Review of Condensed Matter Physics}\ }\textbf {\bibinfo {volume}
  {1}},\ \bibinfo {pages} {129} (\bibinfo {year} {2010})}\BibitemShut {NoStop}%
\bibitem [{\citenamefont {Anderson}(1987)}]{ande87}%
  \BibitemOpen
  \bibfield  {author} {\bibinfo {author} {\bibfnamefont {P.~W.}\ \bibnamefont
  {Anderson}},\ }\href@noop {} {\bibfield  {journal} {\bibinfo  {journal}
  {Science}\ }\textbf {\bibinfo {volume} {235}},\ \bibinfo {pages} {1196}
  (\bibinfo {year} {1987})}\BibitemShut {NoStop}%
\bibitem [{\citenamefont {Lee}\ \emph {et~al.}(2006)\citenamefont {Lee},
  \citenamefont {Nagaosa},\ and\ \citenamefont {Wen}}]{lee06hightc}%
  \BibitemOpen
  \bibfield  {author} {\bibinfo {author} {\bibfnamefont {P.~A.}\ \bibnamefont
  {Lee}}, \bibinfo {author} {\bibfnamefont {N.}~\bibnamefont {Nagaosa}}, \ and\
  \bibinfo {author} {\bibfnamefont {X.-G.}\ \bibnamefont {Wen}},\ }\href@noop
  {} {\bibfield  {journal} {\bibinfo  {journal} {Rev. Mod. Phys.}\ }\textbf
  {\bibinfo {volume} {78}},\ \bibinfo {pages} {17} (\bibinfo {year}
  {2006})}\BibitemShut {NoStop}%
\bibitem [{\citenamefont {J\"ordens}\ \emph {et~al.}(2008)\citenamefont
  {J\"ordens}, \citenamefont {Strohmaier}, \citenamefont {G\"unter},
  \citenamefont {Moritz},\ and\ \citenamefont {Esslinger}}]{jord08}%
  \BibitemOpen
  \bibfield  {author} {\bibinfo {author} {\bibfnamefont {R.}~\bibnamefont
  {J\"ordens}}, \bibinfo {author} {\bibfnamefont {N.}~\bibnamefont
  {Strohmaier}}, \bibinfo {author} {\bibfnamefont {K.}~\bibnamefont
  {G\"unter}}, \bibinfo {author} {\bibfnamefont {H.}~\bibnamefont {Moritz}}, \
  and\ \bibinfo {author} {\bibfnamefont {T.}~\bibnamefont {Esslinger}},\
  }\href@noop {} {\bibfield  {journal} {\bibinfo  {journal} {Nature}\ }\textbf
  {\bibinfo {volume} {455}},\ \bibinfo {pages} {204} (\bibinfo {year}
  {2008})}\BibitemShut {NoStop}%
\bibitem [{\citenamefont {Schneider}\ \emph {et~al.}(2008)\citenamefont
  {Schneider}, \citenamefont {Hackerm\"uller}, \citenamefont {Will},
  \citenamefont {Best}, \citenamefont {Bloch}, \citenamefont {Costi},
  \citenamefont {Helmes}, \citenamefont {Rasch},\ and\ \citenamefont
  {Rosch}}]{schn08}%
  \BibitemOpen
  \bibfield  {author} {\bibinfo {author} {\bibfnamefont {U.}~\bibnamefont
  {Schneider}}, \bibinfo {author} {\bibfnamefont {L.}~\bibnamefont
  {Hackerm\"uller}}, \bibinfo {author} {\bibfnamefont {S.}~\bibnamefont
  {Will}}, \bibinfo {author} {\bibfnamefont {T.}~\bibnamefont {Best}}, \bibinfo
  {author} {\bibfnamefont {I.}~\bibnamefont {Bloch}}, \bibinfo {author}
  {\bibfnamefont {T.~A.}\ \bibnamefont {Costi}}, \bibinfo {author}
  {\bibfnamefont {R.~W.}\ \bibnamefont {Helmes}}, \bibinfo {author}
  {\bibfnamefont {D.}~\bibnamefont {Rasch}}, \ and\ \bibinfo {author}
  {\bibfnamefont {A.}~\bibnamefont {Rosch}},\ }\href@noop {} {\bibfield
  {journal} {\bibinfo  {journal} {Science}\ }\textbf {\bibinfo {volume}
  {322}},\ \bibinfo {pages} {1520} (\bibinfo {year} {2008})}\BibitemShut
  {NoStop}%
\bibitem [{\citenamefont {Taie}\ \emph {et~al.}(2012)\citenamefont {Taie},
  \citenamefont {Yamazaki}, \citenamefont {Sugawa},\ and\ \citenamefont
  {Takahashi}}]{taie2012}%
  \BibitemOpen
  \bibfield  {author} {\bibinfo {author} {\bibfnamefont {S.}~\bibnamefont
  {Taie}}, \bibinfo {author} {\bibfnamefont {R.}~\bibnamefont {Yamazaki}},
  \bibinfo {author} {\bibfnamefont {S.}~\bibnamefont {Sugawa}}, \ and\ \bibinfo
  {author} {\bibfnamefont {Y.}~\bibnamefont {Takahashi}},\ }\href@noop {}
  {\bibfield  {journal} {\bibinfo  {journal} {Nat. Phys.}\ }\textbf {\bibinfo
  {volume} {8}},\ \bibinfo {pages} {825} (\bibinfo {year} {2012})}\BibitemShut
  {NoStop}%
\bibitem [{\citenamefont {Duarte}\ \emph {et~al.}(2015)\citenamefont {Duarte},
  \citenamefont {Hart}, \citenamefont {Yang}, \citenamefont {Liu},
  \citenamefont {Paiva}, \citenamefont {Khatami}, \citenamefont {Scalettar},
  \citenamefont {Trivedi},\ and\ \citenamefont {Hulet}}]{duarte2015}%
  \BibitemOpen
  \bibfield  {author} {\bibinfo {author} {\bibfnamefont {P.~M.}\ \bibnamefont
  {Duarte}}, \bibinfo {author} {\bibfnamefont {R.~A.}\ \bibnamefont {Hart}},
  \bibinfo {author} {\bibfnamefont {T.-L.}\ \bibnamefont {Yang}}, \bibinfo
  {author} {\bibfnamefont {X.}~\bibnamefont {Liu}}, \bibinfo {author}
  {\bibfnamefont {T.}~\bibnamefont {Paiva}}, \bibinfo {author} {\bibfnamefont
  {E.}~\bibnamefont {Khatami}}, \bibinfo {author} {\bibfnamefont {R.~T.}\
  \bibnamefont {Scalettar}}, \bibinfo {author} {\bibfnamefont {N.}~\bibnamefont
  {Trivedi}}, \ and\ \bibinfo {author} {\bibfnamefont {R.~G.}\ \bibnamefont
  {Hulet}},\ }\href@noop {} {\bibfield  {journal} {\bibinfo  {journal} {Phys.
  Rev. Lett.}\ }\textbf {\bibinfo {volume} {114}},\ \bibinfo {pages} {070403}
  (\bibinfo {year} {2015})}\BibitemShut {NoStop}%
\bibitem [{\citenamefont {Greif}\ \emph {et~al.}(2013)\citenamefont {Greif},
  \citenamefont {Uehlinger}, \citenamefont {Jotzu}, \citenamefont {Tarruell},\
  and\ \citenamefont {Esslinger}}]{Greif2013Magnetism}%
  \BibitemOpen
  \bibfield  {author} {\bibinfo {author} {\bibfnamefont {D.}~\bibnamefont
  {Greif}}, \bibinfo {author} {\bibfnamefont {T.}~\bibnamefont {Uehlinger}},
  \bibinfo {author} {\bibfnamefont {G.}~\bibnamefont {Jotzu}}, \bibinfo
  {author} {\bibfnamefont {L.}~\bibnamefont {Tarruell}}, \ and\ \bibinfo
  {author} {\bibfnamefont {T.}~\bibnamefont {Esslinger}},\ }\href@noop {}
  {\bibfield  {journal} {\bibinfo  {journal} {Science}\ }\textbf {\bibinfo
  {volume} {340}},\ \bibinfo {pages} {1307} (\bibinfo {year}
  {2013})}\BibitemShut {NoStop}%
\bibitem [{\citenamefont {Hart}\ \emph {et~al.}(2015)\citenamefont {Hart},
  \citenamefont {Duarte}, \citenamefont {Yang}, \citenamefont {Liu},
  \citenamefont {Paiva}, \citenamefont {Khatami}, \citenamefont {Scalettar},
  \citenamefont {Trivedi}, \citenamefont {Huse},\ and\ \citenamefont
  {Hulet}}]{Hart2015FermiHubbard}%
  \BibitemOpen
  \bibfield  {author} {\bibinfo {author} {\bibfnamefont {R.~A.}\ \bibnamefont
  {Hart}}, \bibinfo {author} {\bibfnamefont {P.~M.}\ \bibnamefont {Duarte}},
  \bibinfo {author} {\bibfnamefont {T.-L.}\ \bibnamefont {Yang}}, \bibinfo
  {author} {\bibfnamefont {X.}~\bibnamefont {Liu}}, \bibinfo {author}
  {\bibfnamefont {T.}~\bibnamefont {Paiva}}, \bibinfo {author} {\bibfnamefont
  {E.}~\bibnamefont {Khatami}}, \bibinfo {author} {\bibfnamefont {R.~T.}\
  \bibnamefont {Scalettar}}, \bibinfo {author} {\bibfnamefont {N.}~\bibnamefont
  {Trivedi}}, \bibinfo {author} {\bibfnamefont {D.~A.}\ \bibnamefont {Huse}}, \
  and\ \bibinfo {author} {\bibfnamefont {R.~G.}\ \bibnamefont {Hulet}},\
  }\href@noop {} {\bibfield  {journal} {\bibinfo  {journal} {Nature}\ }\textbf
  {\bibinfo {volume} {519}},\ \bibinfo {pages} {211} (\bibinfo {year}
  {2015})}\BibitemShut {NoStop}%
\bibitem [{\citenamefont {Greif}\ \emph {et~al.}(2015)\citenamefont {Greif},
  \citenamefont {Jotzu}, \citenamefont {Messer}, \citenamefont {Desbuquois},\
  and\ \citenamefont {Esslinger}}]{Greif2015}%
  \BibitemOpen
  \bibfield  {author} {\bibinfo {author} {\bibfnamefont {D.}~\bibnamefont
  {Greif}}, \bibinfo {author} {\bibfnamefont {G.}~\bibnamefont {Jotzu}},
  \bibinfo {author} {\bibfnamefont {M.}~\bibnamefont {Messer}}, \bibinfo
  {author} {\bibfnamefont {R.}~\bibnamefont {Desbuquois}}, \ and\ \bibinfo
  {author} {\bibfnamefont {T.}~\bibnamefont {Esslinger}},\ }\href@noop {}
  {\bibfield  {journal} {\bibinfo  {journal} {Phys. Rev. Lett.}\ }\textbf
  {\bibinfo {volume} {115}},\ \bibinfo {pages} {260401} (\bibinfo {year}
  {2015})}\BibitemShut {NoStop}%
\bibitem [{\citenamefont {Cocchi}\ \emph {et~al.}(2016)\citenamefont {Cocchi},
  \citenamefont {Miller}, \citenamefont {Drewes}, \citenamefont {Koschorreck},
  \citenamefont {Pertot}, \citenamefont {Brennecke},\ and\ \citenamefont
  {K\"ohl}}]{Cocchi2015}%
  \BibitemOpen
  \bibfield  {author} {\bibinfo {author} {\bibfnamefont {E.}~\bibnamefont
  {Cocchi}}, \bibinfo {author} {\bibfnamefont {L.~A.}\ \bibnamefont {Miller}},
  \bibinfo {author} {\bibfnamefont {J.~H.}\ \bibnamefont {Drewes}}, \bibinfo
  {author} {\bibfnamefont {M.}~\bibnamefont {Koschorreck}}, \bibinfo {author}
  {\bibfnamefont {D.}~\bibnamefont {Pertot}}, \bibinfo {author} {\bibfnamefont
  {F.}~\bibnamefont {Brennecke}}, \ and\ \bibinfo {author} {\bibfnamefont
  {M.}~\bibnamefont {K\"ohl}},\ }\href@noop {} {\bibfield  {journal} {\bibinfo
  {journal} {Phys. Rev. Lett.}\ }\textbf {\bibinfo {volume} {116}},\ \bibinfo
  {pages} {175301} (\bibinfo {year} {2016})}\BibitemShut {NoStop}%
\bibitem [{\citenamefont {Hofrichter}\ \emph {et~al.}(2015)\citenamefont
  {Hofrichter}, \citenamefont {Riegger}, \citenamefont {Scazza}, \citenamefont
  {H\"ofer}, \citenamefont {Fernandes}, \citenamefont {Bloch},\ and\
  \citenamefont {F\"olling}}]{hofrichter2015}%
  \BibitemOpen
  \bibfield  {author} {\bibinfo {author} {\bibfnamefont {C.}~\bibnamefont
  {Hofrichter}}, \bibinfo {author} {\bibfnamefont {L.}~\bibnamefont {Riegger}},
  \bibinfo {author} {\bibfnamefont {F.}~\bibnamefont {Scazza}}, \bibinfo
  {author} {\bibfnamefont {M.}~\bibnamefont {H\"ofer}}, \bibinfo {author}
  {\bibfnamefont {D.~R.}\ \bibnamefont {Fernandes}}, \bibinfo {author}
  {\bibfnamefont {I.}~\bibnamefont {Bloch}}, \ and\ \bibinfo {author}
  {\bibfnamefont {S.}~\bibnamefont {F\"olling}},\ }\href@noop {} {\bibfield
  {journal} {\bibinfo  {journal} {preprint arXiv:1511.07287}\ } (\bibinfo
  {year} {2015})}\BibitemShut {NoStop}%
\bibitem [{\citenamefont {Bakr}\ \emph {et~al.}(2010)\citenamefont {Bakr},
  \citenamefont {Peng}, \citenamefont {Tai}, \citenamefont {Ma}, \citenamefont
  {Simon}, \citenamefont {Gillen}, \citenamefont {F\"olling}, \citenamefont
  {Pollet},\ and\ \citenamefont {Greiner}}]{bakr2010MottInsulator}%
  \BibitemOpen
  \bibfield  {author} {\bibinfo {author} {\bibfnamefont {W.~S.}\ \bibnamefont
  {Bakr}}, \bibinfo {author} {\bibfnamefont {A.}~\bibnamefont {Peng}}, \bibinfo
  {author} {\bibfnamefont {M.~E.}\ \bibnamefont {Tai}}, \bibinfo {author}
  {\bibfnamefont {R.}~\bibnamefont {Ma}}, \bibinfo {author} {\bibfnamefont
  {J.}~\bibnamefont {Simon}}, \bibinfo {author} {\bibfnamefont {J.~I.}\
  \bibnamefont {Gillen}}, \bibinfo {author} {\bibfnamefont {S.}~\bibnamefont
  {F\"olling}}, \bibinfo {author} {\bibfnamefont {L.}~\bibnamefont {Pollet}}, \
  and\ \bibinfo {author} {\bibfnamefont {M.}~\bibnamefont {Greiner}},\
  }\href@noop {} {\bibfield  {journal} {\bibinfo  {journal} {Science}\ }\textbf
  {\bibinfo {volume} {329}},\ \bibinfo {pages} {547} (\bibinfo {year}
  {2010})}\BibitemShut {NoStop}%
\bibitem [{\citenamefont {Sherson}\ \emph {et~al.}(2010)\citenamefont
  {Sherson}, \citenamefont {Weitenberg}, \citenamefont {Endres}, \citenamefont
  {Cheneau}, \citenamefont {Bloch},\ and\ \citenamefont
  {Kuhr}}]{sherson2010microscope}%
  \BibitemOpen
  \bibfield  {author} {\bibinfo {author} {\bibfnamefont {J.~F.}\ \bibnamefont
  {Sherson}}, \bibinfo {author} {\bibfnamefont {C.}~\bibnamefont {Weitenberg}},
  \bibinfo {author} {\bibfnamefont {M.}~\bibnamefont {Endres}}, \bibinfo
  {author} {\bibfnamefont {M.}~\bibnamefont {Cheneau}}, \bibinfo {author}
  {\bibfnamefont {I.}~\bibnamefont {Bloch}}, \ and\ \bibinfo {author}
  {\bibfnamefont {S.}~\bibnamefont {Kuhr}},\ }\href@noop {} {\bibfield
  {journal} {\bibinfo  {journal} {Nature}\ }\textbf {\bibinfo {volume} {467}},\
  \bibinfo {pages} {68} (\bibinfo {year} {2010})}\BibitemShut {NoStop}%
\bibitem [{\citenamefont {Endres}\ \emph {et~al.}(2011)\citenamefont {Endres},
  \citenamefont {Cheneau}, \citenamefont {Fukuhara}, \citenamefont
  {Weitenberg}, \citenamefont {Schau\ss}, \citenamefont {Gross}, \citenamefont
  {Mazza}, \citenamefont {Banuls}, \citenamefont {Pollet}, \citenamefont
  {Bloch},\ and\ \citenamefont {Kuhr}}]{Endres2011stringorder}%
  \BibitemOpen
  \bibfield  {author} {\bibinfo {author} {\bibfnamefont {M.}~\bibnamefont
  {Endres}}, \bibinfo {author} {\bibfnamefont {M.}~\bibnamefont {Cheneau}},
  \bibinfo {author} {\bibfnamefont {T.}~\bibnamefont {Fukuhara}}, \bibinfo
  {author} {\bibfnamefont {C.}~\bibnamefont {Weitenberg}}, \bibinfo {author}
  {\bibfnamefont {P.}~\bibnamefont {Schau\ss}}, \bibinfo {author}
  {\bibfnamefont {C.}~\bibnamefont {Gross}}, \bibinfo {author} {\bibfnamefont
  {L.}~\bibnamefont {Mazza}}, \bibinfo {author} {\bibfnamefont {M.~C.}\
  \bibnamefont {Banuls}}, \bibinfo {author} {\bibfnamefont {L.}~\bibnamefont
  {Pollet}}, \bibinfo {author} {\bibfnamefont {I.}~\bibnamefont {Bloch}}, \
  and\ \bibinfo {author} {\bibfnamefont {S.}~\bibnamefont {Kuhr}},\ }\href@noop
  {} {\bibfield  {journal} {\bibinfo  {journal} {Science}\ }\textbf {\bibinfo
  {volume} {334}},\ \bibinfo {pages} {200} (\bibinfo {year}
  {2011})}\BibitemShut {NoStop}%
\bibitem [{\citenamefont {Cheneau}\ \emph {et~al.}(2012)\citenamefont
  {Cheneau}, \citenamefont {Barmettler}, \citenamefont {Poletti}, \citenamefont
  {Endres}, \citenamefont {Schau\ss}, \citenamefont {Fukuhara}, \citenamefont
  {Gross}, \citenamefont {Bloch}, \citenamefont {Kollath},\ and\ \citenamefont
  {Kuhr}}]{Cheneau2012Lightcone}%
  \BibitemOpen
  \bibfield  {author} {\bibinfo {author} {\bibfnamefont {M.}~\bibnamefont
  {Cheneau}}, \bibinfo {author} {\bibfnamefont {P.}~\bibnamefont {Barmettler}},
  \bibinfo {author} {\bibfnamefont {D.}~\bibnamefont {Poletti}}, \bibinfo
  {author} {\bibfnamefont {M.}~\bibnamefont {Endres}}, \bibinfo {author}
  {\bibfnamefont {P.}~\bibnamefont {Schau\ss}}, \bibinfo {author}
  {\bibfnamefont {T.}~\bibnamefont {Fukuhara}}, \bibinfo {author}
  {\bibfnamefont {C.}~\bibnamefont {Gross}}, \bibinfo {author} {\bibfnamefont
  {I.}~\bibnamefont {Bloch}}, \bibinfo {author} {\bibfnamefont
  {C.}~\bibnamefont {Kollath}}, \ and\ \bibinfo {author} {\bibfnamefont
  {S.}~\bibnamefont {Kuhr}},\ }\href@noop {} {\bibfield  {journal} {\bibinfo
  {journal} {Nature}\ }\textbf {\bibinfo {volume} {481}},\ \bibinfo {pages}
  {484} (\bibinfo {year} {2012})}\BibitemShut {NoStop}%
\bibitem [{\citenamefont {Cheuk}\ \emph {et~al.}(2015)\citenamefont {Cheuk},
  \citenamefont {Nichols}, \citenamefont {Okan}, \citenamefont {Gersdorf},
  \citenamefont {Ramasesh}, \citenamefont {Bakr}, \citenamefont {Lompe},\ and\
  \citenamefont {Zwierlein}}]{Cheuk2015}%
  \BibitemOpen
  \bibfield  {author} {\bibinfo {author} {\bibfnamefont {L.~W.}\ \bibnamefont
  {Cheuk}}, \bibinfo {author} {\bibfnamefont {M.~A.}\ \bibnamefont {Nichols}},
  \bibinfo {author} {\bibfnamefont {M.}~\bibnamefont {Okan}}, \bibinfo {author}
  {\bibfnamefont {T.}~\bibnamefont {Gersdorf}}, \bibinfo {author}
  {\bibfnamefont {V.~V.}\ \bibnamefont {Ramasesh}}, \bibinfo {author}
  {\bibfnamefont {W.~S.}\ \bibnamefont {Bakr}}, \bibinfo {author}
  {\bibfnamefont {T.}~\bibnamefont {Lompe}}, \ and\ \bibinfo {author}
  {\bibfnamefont {M.~W.}\ \bibnamefont {Zwierlein}},\ }\href@noop {} {\bibfield
   {journal} {\bibinfo  {journal} {Phys. Rev. Lett.}\ }\textbf {\bibinfo
  {volume} {114}},\ \bibinfo {pages} {193001} (\bibinfo {year}
  {2015})}\BibitemShut {NoStop}%
\bibitem [{\citenamefont {Haller}\ \emph {et~al.}(2015)\citenamefont {Haller},
  \citenamefont {Hudson}, \citenamefont {Kelly}, \citenamefont {Cotta},
  \citenamefont {Peaudecerf}, \citenamefont {Bruce},\ and\ \citenamefont
  {Kuhr}}]{Haller2015}%
  \BibitemOpen
  \bibfield  {author} {\bibinfo {author} {\bibfnamefont {E.}~\bibnamefont
  {Haller}}, \bibinfo {author} {\bibfnamefont {J.}~\bibnamefont {Hudson}},
  \bibinfo {author} {\bibfnamefont {A.}~\bibnamefont {Kelly}}, \bibinfo
  {author} {\bibfnamefont {D.~A.}\ \bibnamefont {Cotta}}, \bibinfo {author}
  {\bibfnamefont {B.}~\bibnamefont {Peaudecerf}}, \bibinfo {author}
  {\bibfnamefont {G.~D.}\ \bibnamefont {Bruce}}, \ and\ \bibinfo {author}
  {\bibfnamefont {S.}~\bibnamefont {Kuhr}},\ }\href@noop {} {\bibfield
  {journal} {\bibinfo  {journal} {Nat. Phys.}\ }\textbf {\bibinfo {volume}
  {11}},\ \bibinfo {pages} {738} (\bibinfo {year} {2015})}\BibitemShut
  {NoStop}%
\bibitem [{\citenamefont {Parsons}\ \emph {et~al.}(2015)\citenamefont
  {Parsons}, \citenamefont {Huber}, \citenamefont {Mazurenko}, \citenamefont
  {Chiu}, \citenamefont {Setiawan}, \citenamefont {Wooley-Brown}, \citenamefont
  {Blatt},\ and\ \citenamefont {Greiner}}]{Parsons2015}%
  \BibitemOpen
  \bibfield  {author} {\bibinfo {author} {\bibfnamefont {M.~F.}\ \bibnamefont
  {Parsons}}, \bibinfo {author} {\bibfnamefont {F.}~\bibnamefont {Huber}},
  \bibinfo {author} {\bibfnamefont {A.}~\bibnamefont {Mazurenko}}, \bibinfo
  {author} {\bibfnamefont {C.~S.}\ \bibnamefont {Chiu}}, \bibinfo {author}
  {\bibfnamefont {W.}~\bibnamefont {Setiawan}}, \bibinfo {author}
  {\bibfnamefont {K.}~\bibnamefont {Wooley-Brown}}, \bibinfo {author}
  {\bibfnamefont {S.}~\bibnamefont {Blatt}}, \ and\ \bibinfo {author}
  {\bibfnamefont {M.}~\bibnamefont {Greiner}},\ }\href@noop {} {\bibfield
  {journal} {\bibinfo  {journal} {Phys. Rev. Lett.}\ }\textbf {\bibinfo
  {volume} {114}},\ \bibinfo {pages} {213002} (\bibinfo {year}
  {2015})}\BibitemShut {NoStop}%
\bibitem [{\citenamefont {Edge}\ \emph {et~al.}(2015)\citenamefont {Edge},
  \citenamefont {Anderson}, \citenamefont {Jervis}, \citenamefont {McKay},
  \citenamefont {Day}, \citenamefont {Trotzky},\ and\ \citenamefont
  {Thywissen}}]{Edge2015}%
  \BibitemOpen
  \bibfield  {author} {\bibinfo {author} {\bibfnamefont {G.~J.~A.}\
  \bibnamefont {Edge}}, \bibinfo {author} {\bibfnamefont {R.}~\bibnamefont
  {Anderson}}, \bibinfo {author} {\bibfnamefont {D.}~\bibnamefont {Jervis}},
  \bibinfo {author} {\bibfnamefont {D.~C.}\ \bibnamefont {McKay}}, \bibinfo
  {author} {\bibfnamefont {R.}~\bibnamefont {Day}}, \bibinfo {author}
  {\bibfnamefont {S.}~\bibnamefont {Trotzky}}, \ and\ \bibinfo {author}
  {\bibfnamefont {J.~H.}\ \bibnamefont {Thywissen}},\ }\href@noop {} {\bibfield
   {journal} {\bibinfo  {journal} {Phys. Rev. A}\ }\textbf {\bibinfo {volume}
  {92}},\ \bibinfo {pages} {063406} (\bibinfo {year} {2015})}\BibitemShut
  {NoStop}%
\bibitem [{\citenamefont {Omran}\ \emph {et~al.}(2015)\citenamefont {Omran},
  \citenamefont {Boll}, \citenamefont {Hilker}, \citenamefont {Kleinlein},
  \citenamefont {Salomon}, \citenamefont {Bloch},\ and\ \citenamefont
  {Gross}}]{Omran2015}%
  \BibitemOpen
  \bibfield  {author} {\bibinfo {author} {\bibfnamefont {A.}~\bibnamefont
  {Omran}}, \bibinfo {author} {\bibfnamefont {M.}~\bibnamefont {Boll}},
  \bibinfo {author} {\bibfnamefont {T.~A.}\ \bibnamefont {Hilker}}, \bibinfo
  {author} {\bibfnamefont {K.}~\bibnamefont {Kleinlein}}, \bibinfo {author}
  {\bibfnamefont {G.}~\bibnamefont {Salomon}}, \bibinfo {author} {\bibfnamefont
  {I.}~\bibnamefont {Bloch}}, \ and\ \bibinfo {author} {\bibfnamefont
  {C.}~\bibnamefont {Gross}},\ }\href@noop {} {\bibfield  {journal} {\bibinfo
  {journal} {Phys. Rev. Lett.}\ }\textbf {\bibinfo {volume} {115}},\ \bibinfo
  {pages} {263001} (\bibinfo {year} {2015})}\BibitemShut {NoStop}%
\bibitem [{\citenamefont {Greif}\ \emph {et~al.}(2016)\citenamefont {Greif},
  \citenamefont {Parsons}, \citenamefont {Mazurenko}, \citenamefont {Chiu},
  \citenamefont {Blatt}, \citenamefont {Huber}, \citenamefont {Ji},\ and\
  \citenamefont {Greiner}}]{Greif2016}%
  \BibitemOpen
  \bibfield  {author} {\bibinfo {author} {\bibfnamefont {D.}~\bibnamefont
  {Greif}}, \bibinfo {author} {\bibfnamefont {M.~F.}\ \bibnamefont {Parsons}},
  \bibinfo {author} {\bibfnamefont {A.}~\bibnamefont {Mazurenko}}, \bibinfo
  {author} {\bibfnamefont {C.~S.}\ \bibnamefont {Chiu}}, \bibinfo {author}
  {\bibfnamefont {S.}~\bibnamefont {Blatt}}, \bibinfo {author} {\bibfnamefont
  {F.}~\bibnamefont {Huber}}, \bibinfo {author} {\bibfnamefont
  {G.}~\bibnamefont {Ji}}, \ and\ \bibinfo {author} {\bibfnamefont
  {M.}~\bibnamefont {Greiner}},\ }\href@noop {} {\bibfield  {journal} {\bibinfo
   {journal} {Science}\ }\textbf {\bibinfo {volume} {351}},\ \bibinfo {pages}
  {953} (\bibinfo {year} {2016})}\BibitemShut {NoStop}%
\bibitem [{\citenamefont {Scarola}\ \emph {et~al.}(2009)\citenamefont
  {Scarola}, \citenamefont {Pollet}, \citenamefont {Oitmaa},\ and\
  \citenamefont {Troyer}}]{Scarola2009}%
  \BibitemOpen
  \bibfield  {author} {\bibinfo {author} {\bibfnamefont {V.~W.}\ \bibnamefont
  {Scarola}}, \bibinfo {author} {\bibfnamefont {L.}~\bibnamefont {Pollet}},
  \bibinfo {author} {\bibfnamefont {J.}~\bibnamefont {Oitmaa}}, \ and\ \bibinfo
  {author} {\bibfnamefont {M.}~\bibnamefont {Troyer}},\ }\href@noop {}
  {\bibfield  {journal} {\bibinfo  {journal} {Phys. Rev. Lett.}\ }\textbf
  {\bibinfo {volume} {102}},\ \bibinfo {pages} {135302} (\bibinfo {year}
  {2009})}\BibitemShut {NoStop}%
\bibitem [{\citenamefont {Paiva}\ \emph {et~al.}(2010)\citenamefont {Paiva},
  \citenamefont {Scalettar}, \citenamefont {Randeria},\ and\ \citenamefont
  {Trivedi}}]{Paiva2010}%
  \BibitemOpen
  \bibfield  {author} {\bibinfo {author} {\bibfnamefont {T.}~\bibnamefont
  {Paiva}}, \bibinfo {author} {\bibfnamefont {R.}~\bibnamefont {Scalettar}},
  \bibinfo {author} {\bibfnamefont {M.}~\bibnamefont {Randeria}}, \ and\
  \bibinfo {author} {\bibfnamefont {N.}~\bibnamefont {Trivedi}},\ }\href@noop
  {} {\bibfield  {journal} {\bibinfo  {journal} {Phys. Rev. Lett.}\ }\textbf
  {\bibinfo {volume} {104}},\ \bibinfo {pages} {066406} (\bibinfo {year}
  {2010})}\BibitemShut {NoStop}%
\bibitem [{\citenamefont {DePue}\ \emph {et~al.}(1999)\citenamefont {DePue},
  \citenamefont {McCormick}, \citenamefont {Winoto}, \citenamefont {Oliver},\
  and\ \citenamefont {Weiss}}]{depu99}%
  \BibitemOpen
  \bibfield  {author} {\bibinfo {author} {\bibfnamefont {M.~T.}\ \bibnamefont
  {DePue}}, \bibinfo {author} {\bibfnamefont {C.}~\bibnamefont {McCormick}},
  \bibinfo {author} {\bibfnamefont {S.~L.}\ \bibnamefont {Winoto}}, \bibinfo
  {author} {\bibfnamefont {S.}~\bibnamefont {Oliver}}, \ and\ \bibinfo {author}
  {\bibfnamefont {D.~S.}\ \bibnamefont {Weiss}},\ }\href@noop {} {\bibfield
  {journal} {\bibinfo  {journal} {Phys. Rev. Lett.}\ }\textbf {\bibinfo
  {volume} {82}},\ \bibinfo {pages} {2262} (\bibinfo {year}
  {1999})}\BibitemShut {NoStop}%
\bibitem [{\citenamefont {Khatami}\ and\ \citenamefont
  {Rigol}(2011)}]{khatami2011}%
  \BibitemOpen
  \bibfield  {author} {\bibinfo {author} {\bibfnamefont {E.}~\bibnamefont
  {Khatami}}\ and\ \bibinfo {author} {\bibfnamefont {M.}~\bibnamefont
  {Rigol}},\ }\href@noop {} {\bibfield  {journal} {\bibinfo  {journal} {Phys.
  Rev. A}\ }\textbf {\bibinfo {volume} {84}},\ \bibinfo {pages} {053611}
  (\bibinfo {year} {2011})}\BibitemShut {NoStop}%
\bibitem [{\citenamefont {Oitmaa}\ \emph {et~al.}(2006)\citenamefont {Oitmaa},
  \citenamefont {Hamer},\ and\ \citenamefont {Zheng}}]{Oitmaa2006}%
  \BibitemOpen
  \bibfield  {author} {\bibinfo {author} {\bibfnamefont {J.}~\bibnamefont
  {Oitmaa}}, \bibinfo {author} {\bibfnamefont {C.}~\bibnamefont {Hamer}}, \
  and\ \bibinfo {author} {\bibfnamefont {W.}~\bibnamefont {Zheng}},\
  }\href@noop {} {\emph {\bibinfo {title} {Series Expansion Methods for
  Strongly Interacting Lattice Models}}}\ (\bibinfo  {publisher} {Cambridge
  University Press},\ \bibinfo {address} {Cambridge, England},\ \bibinfo {year}
  {2006})\BibitemShut {NoStop}%
\bibitem [{\citenamefont {Hirsch}(1985)}]{Hirsch1985}%
  \BibitemOpen
  \bibfield  {author} {\bibinfo {author} {\bibfnamefont {J.~E.}\ \bibnamefont
  {Hirsch}},\ }\href@noop {} {\bibfield  {journal} {\bibinfo  {journal} {Phys.
  Rev. B}\ }\textbf {\bibinfo {volume} {31}},\ \bibinfo {pages} {4403}
  (\bibinfo {year} {1985})}\BibitemShut {NoStop}%
\bibitem [{\citenamefont {Paiva}\ \emph {et~al.}(2001)\citenamefont {Paiva},
  \citenamefont {Scalettar}, \citenamefont {Huscroft},\ and\ \citenamefont
  {McMahan}}]{Paiva2001}%
  \BibitemOpen
  \bibfield  {author} {\bibinfo {author} {\bibfnamefont {T.}~\bibnamefont
  {Paiva}}, \bibinfo {author} {\bibfnamefont {R.~T.}\ \bibnamefont
  {Scalettar}}, \bibinfo {author} {\bibfnamefont {C.}~\bibnamefont {Huscroft}},
  \ and\ \bibinfo {author} {\bibfnamefont {A.~K.}\ \bibnamefont {McMahan}},\
  }\href@noop {} {\bibfield  {journal} {\bibinfo  {journal} {Phys. Rev. B}\
  }\textbf {\bibinfo {volume} {63}},\ \bibinfo {pages} {125116} (\bibinfo
  {year} {2001})}\BibitemShut {NoStop}%
\bibitem [{\citenamefont {LeBlanc}\ and\ \citenamefont
  {Gull}(2013)}]{LeBlanc2013}%
  \BibitemOpen
  \bibfield  {author} {\bibinfo {author} {\bibfnamefont {J.~P.~F.}\
  \bibnamefont {LeBlanc}}\ and\ \bibinfo {author} {\bibfnamefont
  {E.}~\bibnamefont {Gull}},\ }\href@noop {} {\bibfield  {journal} {\bibinfo
  {journal} {Phys. Rev. B}\ }\textbf {\bibinfo {volume} {88}},\ \bibinfo
  {pages} {155108} (\bibinfo {year} {2013})}\BibitemShut {NoStop}%
\bibitem [{\citenamefont {Chiesa}\ \emph {et~al.}(2011)\citenamefont {Chiesa},
  \citenamefont {Varney}, \citenamefont {Rigol},\ and\ \citenamefont
  {Scalettar}}]{Chiesa2011}%
  \BibitemOpen
  \bibfield  {author} {\bibinfo {author} {\bibfnamefont {S.}~\bibnamefont
  {Chiesa}}, \bibinfo {author} {\bibfnamefont {C.~N.}\ \bibnamefont {Varney}},
  \bibinfo {author} {\bibfnamefont {M.}~\bibnamefont {Rigol}}, \ and\ \bibinfo
  {author} {\bibfnamefont {R.~T.}\ \bibnamefont {Scalettar}},\ }\href@noop {}
  {\bibfield  {journal} {\bibinfo  {journal} {Phys. Rev. Lett.}\ }\textbf
  {\bibinfo {volume} {106}},\ \bibinfo {pages} {035301} (\bibinfo {year}
  {2011})}\BibitemShut {NoStop}%
\bibitem [{\citenamefont {Weitenberg}\ \emph {et~al.}(2011)\citenamefont
  {Weitenberg}, \citenamefont {Endres}, \citenamefont {Sherson}, \citenamefont
  {Cheneau}, \citenamefont {Schausz}, \citenamefont {Fukuhara}, \citenamefont
  {Bloch},\ and\ \citenamefont {Kuhr}}]{weitenberg2011}%
  \BibitemOpen
  \bibfield  {author} {\bibinfo {author} {\bibfnamefont {C.}~\bibnamefont
  {Weitenberg}}, \bibinfo {author} {\bibfnamefont {M.}~\bibnamefont {Endres}},
  \bibinfo {author} {\bibfnamefont {J.~F.}\ \bibnamefont {Sherson}}, \bibinfo
  {author} {\bibfnamefont {M.}~\bibnamefont {Cheneau}}, \bibinfo {author}
  {\bibfnamefont {P.}~\bibnamefont {Schausz}}, \bibinfo {author} {\bibfnamefont
  {T.}~\bibnamefont {Fukuhara}}, \bibinfo {author} {\bibfnamefont
  {I.}~\bibnamefont {Bloch}}, \ and\ \bibinfo {author} {\bibfnamefont
  {S.}~\bibnamefont {Kuhr}},\ }\href {http://dx.doi.org/10.1038/nature09827}
  {\bibfield  {journal} {\bibinfo  {journal} {Nature}\ }\textbf {\bibinfo
  {volume} {471}},\ \bibinfo {pages} {319} (\bibinfo {year}
  {2011})}\BibitemShut {NoStop}%
\bibitem [{\citenamefont {Park}\ \emph {et~al.}(2015)\citenamefont {Park},
  \citenamefont {Will},\ and\ \citenamefont {Zwierlein}}]{park2015groundstate}%
  \BibitemOpen
  \bibfield  {author} {\bibinfo {author} {\bibfnamefont {J.~W.}\ \bibnamefont
  {Park}}, \bibinfo {author} {\bibfnamefont {S.~A.}\ \bibnamefont {Will}}, \
  and\ \bibinfo {author} {\bibfnamefont {M.~W.}\ \bibnamefont {Zwierlein}},\
  }\href {\doibase 10.1103/PhysRevLett.114.205302} {\bibfield  {journal}
  {\bibinfo  {journal} {Phys. Rev. Lett.}\ }\textbf {\bibinfo {volume} {114}},\
  \bibinfo {pages} {205302} (\bibinfo {year} {2015})}\BibitemShut {NoStop}%
\end{thebibliography}

\pagebreak
\clearpage
\setcounter{equation}{0}
\setcounter{figure}{0}
\renewcommand{\thefigure}{S\arabic{figure}}

\onecolumngrid
\begin{center}
\large{\textbf{Supplemental Material: \\ Observation of 2D fermionic Mott insulators of $^{40}$K with single-site resolution}}
\end{center}
\twocolumngrid

\subsection{Experimental Setup}

To achieve high-resolution imaging, a glass spherical cap is optically contacted to the outside of a vacuum window, while a super-polished substrate is contacted to the inside, as shown in Fig.~S1. These three components (the spherical cap, vacuum window, and substrate) together form a solid immersion lens. When combined with a microscope objective with a numerical aperture (NA) of $0.6$, this allows us to achieve an enhanced NA of $0.87$. The 2D optical lattice is formed by two $1064\,\rm{nm}$ beams propagating along the $x$- and $y$-directions (beams A and B respectively). These beams are reflected from the substrate at an angle of $10.8^{\circ}$ with respect to the substrate surface, and are then retro-reflected. In addition to forming a square lattice in the $x$-$y$ plane with a lattice spacing of $541\,\rm{nm}$, they also form a vertical lattice with a lattice spacing of $2.8\,\mu\rm{m}$ due to the reflection from the substrate. To provide extra confinement during Raman imaging, an additional $1064\,\rm{nm}$ beam (beam C) is turned on, which propagates along the $z$-direction and is retro-reflected off of the substrate. This produces a vertical lattice with a spacing of $532\,\rm{nm}$. An $830\,\rm{nm}$ dimple beam (beam D) with waist $20\,\mu\rm{m}$ is sent through the microscope objective along the $-z$-direction to provide additional radial confinement in the $x$-$y$ plane.

Four additional $1064\,\rm{nm}$ beams are used for transport and preparation of the atoms in a single layer of the vertical lattice. Two beams (E,F) propagating along the $x$- and $y$-directions, respectively, form a crossed optical dipole trap whose trap center can be moved along the $z$-direction using acousto-optical deflectors (AODs). An ``accordion" beam (G) reflects off the substrate surface with a variable angle. Finally, a shallow angle beam (H) propagating along the $x$-direction reflects off the substrate at an angle of $5.8^{\circ}$ with respect to the substrate surface.

\begin{figure}[!]
\centering
\includegraphics[scale=1.0]{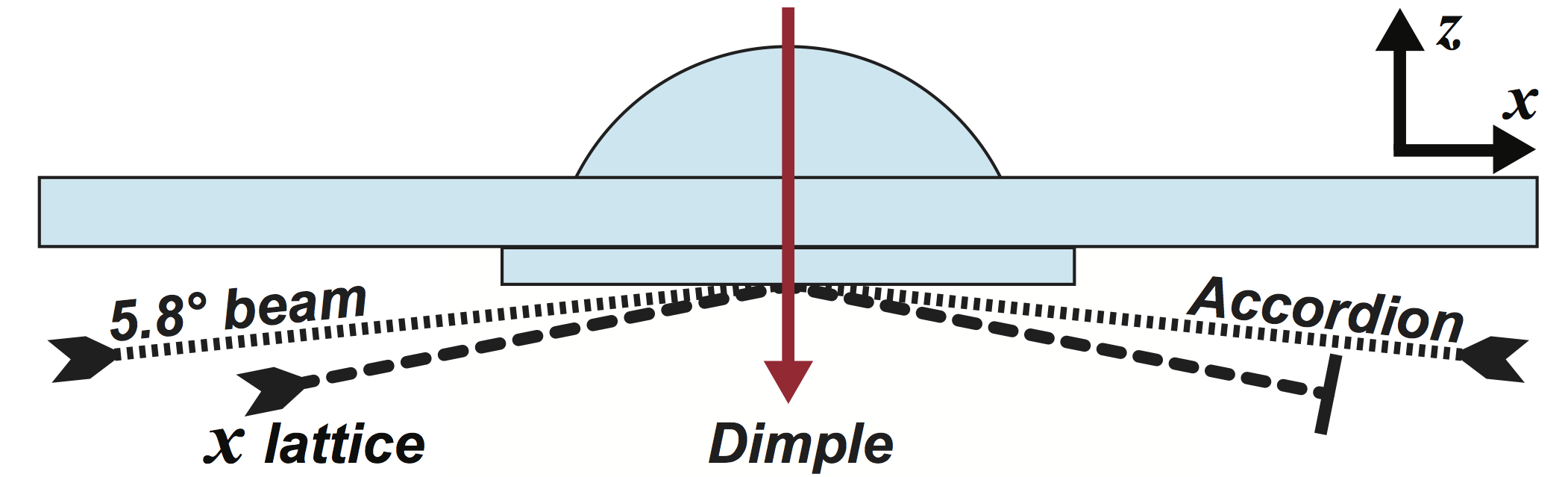}
\caption{High-resolution imaging setup and beam configuration. The solid immersion lens, composed of a glass spherical cap, a vacuum window, and a super-polished substrate is shown. The black dotted line shows the accordion (G) and shallow angle (H) beams, which propagate along the $-x$ and $x$-directions, respectively. The black dashed line shows the $x$-lattice beam (A). The solid red line shows the dimple beam (D). Not shown here are the two dipole trap beams (E,F), the vertical lattice beam (C), and the $y$-lattice beam (B).}

\label{figS1}
\end{figure}

\subsection{Evaporation and Transport Sequence}
The evaporation sequence begins with $^{40}$K atoms in the $\left|F=9/2, m_F=9/2 \ket$ hyperfine state and $^{23}$Na atoms in the $\left| 2,2\ket$ state held in a plugged magnetic trap $2\,\rm{mm}$ below the substrate.  After all of the $^{23}$Na atoms are evaporated with a radio-frequency (RF) sweep on the $\left|2,2 \ket$ $\rightarrow$ $\left|1,1 \ket$ transition, the remaining  $^{40}$K atoms are transferred to a crossed optical dipole trap formed by beams E and F. Subsequently, an equal mixture of $\left|9/2, -9/2 \ket$  and $\left|9/2, -7/2 \ket$ is prepared via RF sweeps. The atoms are further evaporated by lowering the power of the beams. The optical trap is moved over $1.8\,\rm{s}$ to $40\,\mu\rm{m}$ below the substrate by changing the drive frequencies of the AODs. The $x$-dipole trap beam (E) is then ramped down, while the accordion beam (G) is ramped up. Initially, the accordion beam (G) is reflected at $1.2^{\circ}$ off the substrate, forming a vertical lattice with a spacing of $25\,\mu\rm{m}$. The atoms are then loaded into the second layer of this lattice. While the accordion beam (G) is ramped up, the dimple beam (D) is also ramped up. Subsequently, the angle of the accordion beam (G) is increased over $250\,\rm{ms}$ to $5.8^{\circ}$. This compresses the atomic cloud vertically, and transports it to $\sim 7\,\mu\rm{m}$ from the substrate. The accordion beam (G) and the $y$-dipole trap beam (F) are then ramped down while the shallow angle beam (H) is ramped up. The atoms at this stage are thus confined vertically by the shallow angle beam (H) and radially by the dimple beam (D). A final stage of evaporation prior to lattice loading is carried out  by lowering the powers of these two beams.

\subsection{Lattice Ramps}
To load atoms into the lattice, the $x$- and $y$-lattice beams (A,B) are first ramped up over $40\,\rm{ms}$ to $1.5\,E_R$, where $E_R= \frac{\hbar^2}{2m} \frac{\pi^2}{a^2}$, $a$ is the lattice spacing, and $m$ is the mass of a $^{40}$K atom. Subsequently, the lattice depth in the $x$- and $y$-directions is increased to a depth of either $6\,E_R$, $12\,E_R$, or $18\,E_R$. For the first two values, the lattice depth is increased over $100\,\rm{ms}$, while for the deepest lattice depth, the ramp time is $350\,\rm{ms}$.  The radial confinement, controlled by the power of the dimple beam (D), is either decreased or increased during the first $100\,\rm{ms}$ of this ramp. To freeze the atoms, the $x$-$y$ lattices are first ramped over $2\,\rm{ms}$ to $\sim 100\,E_R$. The two lattices are then further increased to $\sim 1000\,E_R$ over the next $40\,\rm{ms}$, while the radial confining beam (D) is ramped down. The shallow angle beam (H) is also ramped down during this time, while the lattice beam along the $z$-direction (C) is ramped up to $\sim 1000\,E_R$. Raman imaging begins $5\,\rm{ms}$ after this ramp.

\subsection{Determination of Hubbard parameters}
To determine the interaction parameter $U$ at $12\,E_R$ and $18\,E_R$, we perform modulation spectroscopy on Mott insulators. When the modulation frequency equals $U/h$, doublon-hole pairs are created, which appear as empty sites due to parity projection. The loss spectra at $12\,E_R$ and $18\,E_R$ are shown in Fig.~S2. The interaction parameter at $6\,E_R$ is obtained via a tight-binding calculation using the measured lattice depths.

\begin{figure}[!]
\centering
\includegraphics[scale=1.1]{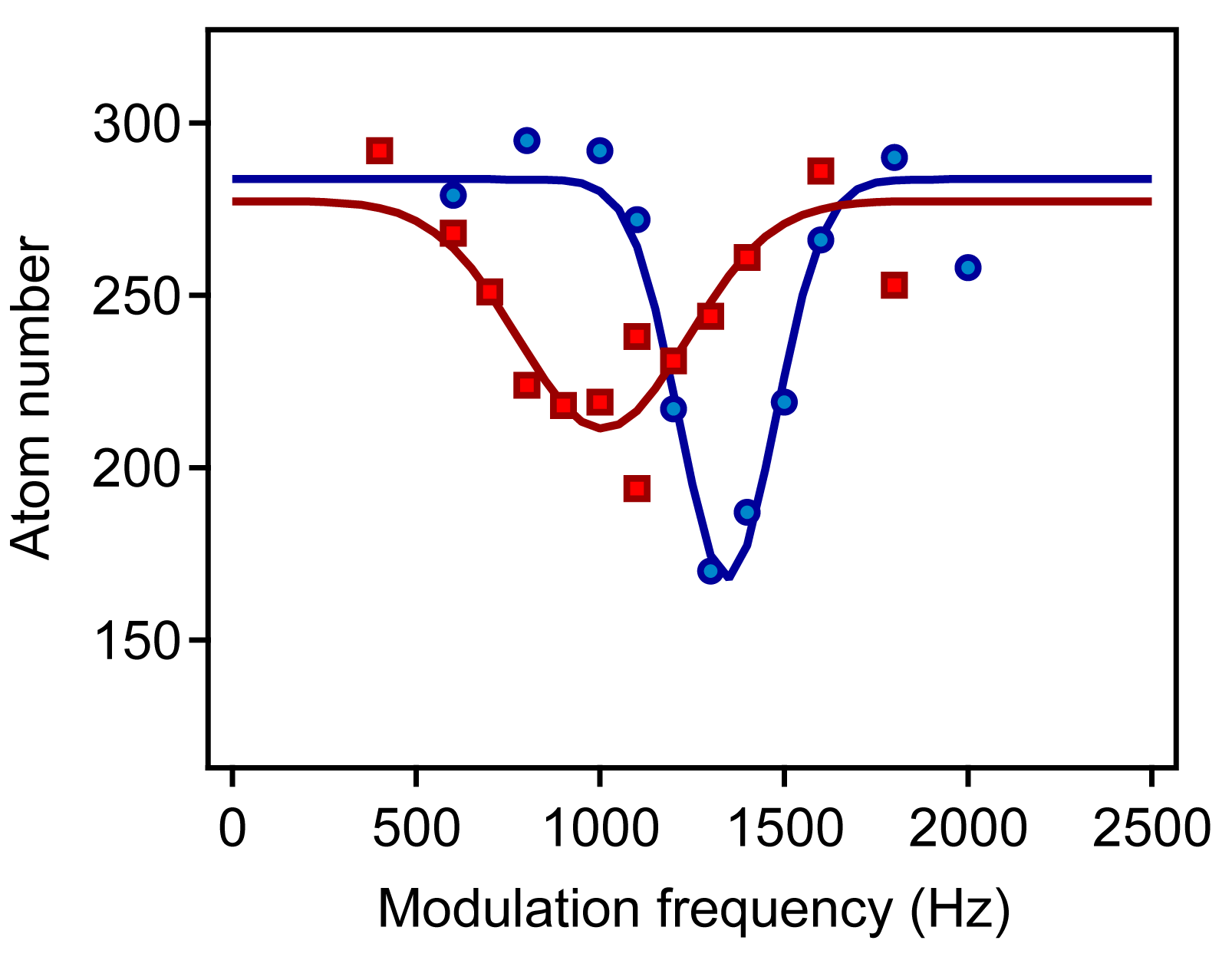}
\caption{Measurement of the Hubbard interaction parameter $U$ via lattice modulation spectroscopy. The blue circles (red squares) show the detected atom number as a function of modulation frequency at a lattice depth of $18\,E_R$ ($12\,E_R$). One lattice beam is modulated with $10\,\%$ ($5\,\%$) amplitude for 1000 (500) cycles of the modulation. The solid blue and red lines are Gaussian fits to the data. The measured interaction parameters for the two lattice depths are $U/h = 1350(50)\,\rm{Hz}$ and $U/h = 1007(40)\,\rm{Hz}$ for $18\,E_R$ and $12\,E_R$, respectively.}
\label{figS2}
\end{figure}

We also perform modulation spectroscopy to determine the lattice depths. The loss features corresponding to ground-to-first-band and ground-to-second-band transitions are identified. To ensure that the system is in the 2D and single-band regime, we measure the vertical trapping frequency also through modulation spectroscopy. We find that for the lattice depths used, the vertical trapping frequency $\omega_z$ ranges from $2\pi \times 4\,\rm{kHz}$ to $2\pi \times 8\,\rm{kHz}$. This corresponds to $\mu/\hbar \omega_z < 0.15$, where $\mu$ is the chemical potential.

\end{document}